\documentclass[fleqn,10pt]{wlscirep}
\usepackage[utf8]{inputenc}
\usepackage[T1]{fontenc}
\usepackage[normalem]{ulem}
\usepackage{siunitx}
\usepackage{tikz}
\usepackage{graphicx} 
\usepackage[table]{xcolor}
\usepackage{array}
\usepackage{float}
\usepackage{amsmath,amssymb,amsfonts}
\usepackage{makecell}
\usepackage{url}
\usepackage{comment}
\usepackage{subcaption}

\newcommand{\blueHL}[1]{\textcolor{blue}{#1}}

\newcommand{\arch}{HOIM}

\title{Towards Higher Order Oscillatory Ising Machines}

\author[1,*]{Nafisa Sadaf Prova}
\author[1]{Hüsrev Cılasun}
\author[1]{Abhimanyu Kumar}
\author[1]{Ahmet Efe}
\author[1]{Sachin S. Sapatnekar}
\author[1]{Ulya R. Karpuzcu}

\affil[1]{University of Minnesota, Minneapolis, United States}
\affil[*]{Corresponding author, prova026@umn.edu}

\begin{abstract}
Ising machines as hardware solvers of combinatorial optimization problems (COPs) 
can efficiently explore large solution spaces due to their inherent parallelism and 
physics-based dynamics.
Many important COPs such as satisfiability (SAT) 
assume arbitrary interactions between problem variables, while most 
Ising machines only support
pairwise (second-order) interactions. This necessitates translation of higher-order interactions to pairwise, which typically results in extra variables not corresponding to problem variables, and a larger problem for the Ising machine to solve than the original problem. This in turn can significantly increase time-to-solution and/or degrade solution accuracy. In this paper, considering a representative CMOS-compatible class of Ising machines, we propose a practical design to enable direct hardware support for higher order interactions. By minimizing the overhead of problem translation and mapping, 
our design can result in up to 4 $\times$ lower time-to-solution without compromising solution accuracy.

\end{abstract}

\begin{document}

\maketitle

\section{Introduction} \label{sec:Introduction}

Solutions to combinatorial optimization problems (COP) 
come from a finite discrete search space, and
minimize or maximize an objective function
under a given set of constraints. 
COPs are amongst the hardest optimization problems;
most of the COPs belong to the computational complexity classes NP-complete or NP-hard. At the same time, problem sizes of practical interest prohibit exhaustive search in the discrete solution space. As a result,  classical solvers typically rely on approximations and cannot always guarantee to find solutions. 

Ising machines as COP solvers can efficiently explore large solution spaces by directly leveraging the inherent parallelism and natural dynamic behavior of Ising-model compliant physical systems which stabilize at minimum energy states.
The Ising model, originally developed in statistical mechanics to describe ferromagnetism, 
represents each solution as a binary vector, where each element maps to a spin state in such a way that the extremes of the objective function match the minimum energy of the system.
As the system naturally favors spin configurations corresponding to the minimum energy states, it essentially solves a COP by converging on a minimum energy state upon perturbation.


In their most basic and most common form, Ising machines support
pairwise (second-order) interactions between spins. 
%
However, many real-world COPs from
computational biology~\cite{PhysRevResearch.4.023062}, machine learning~\cite{Laydevant2024} or financial modeling~\cite{PhysRevE.103.062130} involve 
higher-order interactions between three or more variables.
Such higher-order interactions 
are typically
approximated or transformed into equivalent pairwise interactions through a process known as \textit{quadratization},
by introducing auxiliary variables which do not correspond to problem variables. Quadratization can also increase the range of core model parameters -- such as interaction coefficients quantifying the strength of interactions -- to exceed physical representation capabilities of the targeted Ising machines.
This results in a larger, and often more complex, problem for the Ising machine to solve than the original problem, which in turn can significantly increase time-to-solution and/or degrade solution accuracy.

Depending on the underlying physical system, Ising machines come in different flavors.
In this paper, we confine our study to {\em Oscillatory Ising Machines (OIM)}~\cite{Wang2021,Chou2019,lo2023ising}, 
a practical class of Ising machines based on conventional technologies and operating at room temperature.  An OIM is a network of coupled oscillators
%
engineered to have their interactions encode the constraints and objectives of a given COP.
As the system evolves, the collective dynamics of the oscillators guide the OIM toward a stable configuration, ideally corresponding to an optimal or near-optimal solution. 
Each oscillator maps to a spin, where two distinct values of the oscillator phase encode the two distinct spin states (spin-up, +1 or spin-down, -1). To determine whether 
the oscillator phase at a given point in time encodes spin-up or spin-down, we need a \textit{reference} --
a pre-specified fixed phase 
against which each oscillator’s phase is compared. 
OIMs typically use a set of phase-locked reference oscillators to this end.

Dedicated hardware elements called {\em couplers} implement pairwise spin interactions such that oscillators can influence each other's phase.
A key challenge for augmenting OIM couplers with support for
higher-order interactions is distributing the reference phase to each coupler across the network, which 
requires increasingly complex circuitry and interconnects at scale.
As a result, most representative OIMs from the literature 
only support pairwise coupling -- and must rely on problem translation to be able to solve COPs featuring higher-order interactions, which incurs forbidding overheads at scale. 
We will refer to these designs as  
Pairwise Oscillatory Ising Machines (POIM). 

In this paper 
 we propose a practical design to enable Higher-Order Ising Machines (\arch) which 
 natively support higher order interactions.
Specifically we introduce a higher-order coupler design that does not need a reference to operate. 
In mapping higher-order COPs, these higher-order couplers facilitate seamless problem translation which prevents the introduction of ancillary variables or the expansion of interaction coefficient ranges beyond machine capability.

\section{Background} \label{sec:Background}
\textbf{{Ising Model:}}
As a mathematical abstraction, in its most basic form, {Ising Model}~\cite{1986JPhA...19.1605F,Barahona_1982,Farhi_2001}
describes any physical system composed of pairwise interacting discrete elements such as spins~\cite{ising1925beitrag}.
A spin can be either aligned or anti-aligned with an external magnetic field. 
A Hamiltonian function is used to describe the energy of an $n$-spin system with $s=\left[ s_1, s_2, ..., s_n \right]$, 
as:
$$H(s) = -\sum_{<i\neq j>} J_{ij} s_i s_j -\sum_i h_i s_i ~~~ \text{where}~~~i,j \in [1,2, ..., n]$$

\noindent $s$ is the vector capturing spin states, where spin $s_i$
can be either $-1$ or $+1$ (which encode its alignment). 
The spins 
reside in a finite-dimensional lattice, often visualized as a graph.
$h_i$ captures the strength of the {\em local} field at $s_i$; 
$J_{ij}$, of the {\em interaction} or {\em coupling} between neighboring spins $s_i$ and $s_j$. $h_i$ and $J_{ij}$ are real-valued constants.

The system evolves toward the ground state $s=arg min_s H(s)$,
which minimizes the energy function $H({s})$ at equilibrium. 
The interaction between neighboring spins depends on the sign of the coupling coefficient $J_{ij}$. Specifically, if $J_{ij}$ is positive, 
neighboring spins tend to align such that $s_i s_j = +1$; if $J_{ij}$ is negative, 
to anti-align, such that $s_i s_j = -1$. 
Both cases result in a negative term $- |J_{ij}|$ to reduce 
the total energy $H({s})$. 

The Ising model is mathematically equivalent to Quadratic Unconstrained Binary Optimization (QUBO)~\cite{QUBO} characterized by 
$$H({x})={x}^T Q {x}~~\text{where}~~ x_i = (s_i + 1)/2$$ 
$$h_i = Q_{ii}/2 + \sum_{j=1}^n (Q_{ij} + Q_{ji})/4 ~~\text{and}~~ J_{ij}=Q_{ij}/4$$
for ${x} = [x_1, \cdots , x_n]^T \in \{ 0, 1 \}^n$ and the QUBO matrix $Q \in \mathbb{R}^{n \times n}$ for $n$ spins.

As COPs grow in size, exhaustive search -- i.e., evaluating $H(s)$ for all possible configurations of $s$ within the search space termed the {\em energy landscape} -- becomes computationally infeasible. 
%
Ising model based software solvers typically utilize probabilistic methods such as simulated annealing~\cite{sa}, which involves iteratively flipping selected spin states to explore the energy landscape according to predefined convergence criteria. 
On the other hand, Ising model based hardware solvers can be broadly categorized into two classes. The first class~\cite{takemoto20192,yamamoto2020statica,aramon2019physics,goto2021high,tatsumura2021scaling} treats the Ising model purely as a mathematical abstraction, mirroring software-based solvers but with hardware acceleration. 
The second class, which we refer to as {\em Ising machines}, implements physical systems that inherently follow Ising model dynamics~\cite{dutta2021ising,moy20221,Lo2023,inagaki2016coherent}. Oscillatory Ising Machines (OIM) fall into this category, where underlying physics governs
the evolution of spin states over time.

\noindent {\textbf{Problem Formulation}:} Formulating a COP using the Ising model involves
representing problem variables as 
spins and encoding constraints through 
pairwise coupling strengths $J_{ij}$.
Ising formulations for a wide range of NP-complete and NP-hard problems -- including all 21 of Karp’s -- 
exist~\cite{Lucas_2014}. The goal is to construct an energy function (Hamiltonian) 
with the 
minimum-energy state corresponding to the optimal solution of the original problem. 
Then, an Ising machine
can, with high probability, solve complex COPs as the system relaxes into its ground state, 
even under the influence of 
noise~\cite{isingBasicsDWave}.

For a given optimization problem, multiple Ising or QUBO formulations may be possible. Both Ising and QUBO models inherently represent pairwise interactions between spins, and by extension, between problem variables. However, many COPs involve interactions among more than just two variables. A notable example is SAT(isfiability),
a foundational problem in computer science
\cite{cormen2022introduction}, which was the first problem proven to be NP-complete and which served as the starting point for the theory of NP-completeness. Any other NP-complete problem can be transformed into SAT using a polynomial-time reduction~\cite{karp2010reducibility}.
Accordingly, we use SAT to benchmark the proposed designs.

SAT 
involves
finding 
an assignment of truth values (true or false) to $n$ Boolean variables $X = \{x_1, x_2, \ldots, x_n\}$ that satisfies a given Boolean formula $f(x_1, x_2, \ldots, x_n)$ -- typically in Conjunctive Normal Form (CNF). Each clause $C = \{C_1, C_2, \ldots, C_m\}$ in the CNF formula 
expresses higher-order interactions between problem variables by correlating literals (Boolean variables or their negations) using Boolean OR ($\lor$), Boolean AND ($\land$), and Boolean negation ($\neg$):
%
$$f(x_1, \cdots, x_n) = C_1 \land C_2 \land \cdots \land C_m ~~\text{with}~~ X = \{x_1, \cdots, x_n\}$$
%
Generally, clauses in a k-SAT problem can contain at most k literals. 3-SAT (k = 3) is most commonly studied, and any k-SAT instance with k $>$ 3 can be efficiently reduced to an equivalent 3-SAT problem in polynomial time.
In 3SAT, 
each clause $C_i = l_1 \lor l_2 \lor l_3$ 
can have 
at most 3 literals $l_1, l_2, l_3 \subset X \cup \neg X$. 

Considering a generic 3-SAT clause,
translating the cubic (third-order) interaction among the problem variables 
into the equivalent quadratic (second-order or pairwise) expression -- {quadratization},
typically results in new binary variables (ancillaries) to express the cubic interaction 
through extra 
terms which penalize unsatisfying assignments~\cite{cilasun2024sat}.
How such extra terms look like and how many ancillary variables each necessitates to encode problem-specific correlations 
depend on the mathematical formulation.  

To translate an $n$-variable $m$-clause 3-SAT instance, even the highly compact formulation proposed by Chancellor~\cite{Chancellor2016} introduces an ancillary variable for each clause, resulting in a total of $n$+$m$ variables in the translated problem. Alternatively, in the Integer Linear Programming (ILP)-based formulation proposed by~\cite{cilasun2024sat}, the encoding uses two ancillary variables per clause to model clause Hamiltonians as equality constraints, yielding a total of $n$+2$m$ variables in the translated problem. 
Ancillary
variables ensure that multi-variable logical expressions are correctly transformed into quadratic (pairwise) energy penalties, making the SAT-to-Ising (or equivalently QUBO) mapping both physically implementable and mathematically correct. 

In contrast to these formulations, mapping a SAT instance to HOIMs requires solving a Polynomial Unconstrained Binary Optimization (PUBO)~\cite{nagies2025boostingquantumannealingperformance} problem instead of QUBO -- which we 
will 
refer to as 
Higher-order Unconstrained Binary Optimization (HUBO) throughout the paper, and which eliminates the need for ancillary variables by definition.

\noindent {\textbf{Coupler Design Space:}}
The performance of Ising machines strongly depends on the {\em coupler design} -- i.e., the physical implementation of interaction (or coupling) coefficients ($J_{ij}$). There are numerous ways to translate coupling coefficients into physical spin-to-spin couplings.
%
Most architectures store the coupling coefficients in (SRAM) memory. The difference comes from the mechanism to realize the interaction dynamics.
\textit{Transmission-gate} based couplers are commonly used in OIMs and employ SRAM-programmed transmission-gate stages to electrically couple oscillator phases to become either in-phase or out-of-phase with each other. Transmission gates effectively serve as digital switches in this case. 
Couplers from the OIM design in~\cite{lo2023ising}, as a representative example,
consists of cascaded stages of transmission gates controlled by SRAM bits, enabling programmable integer-valued coupling strengths.
In typical OIMs where oscillator phases represent spin states, however, such couplers can directly map at most quadratic (pairwise) interactions.
In mathematical terms, quadratic couplers correlate two oscillators $i$ and $j$ by adjusting the degree of correlation between their phases according to the 
value of the interaction coefficient $J_{ij}$.
%
Couplers are critical design components, as OIMs can only evolve to a solution based on the interaction between the oscillator phases. In other words, 
coupling dynamics between the oscillator phases guide OIMs toward the global Hamiltonian minimum.

\section{Higher Order Oscillatory Ising Machines} 
\label{sec:hoim}

\subsection{Mathematical Formulation}
\label{sec:Implementation}
We next take a closer look into 3-SAT as an illustrative example, considering the generic 3-SAT clause  \( (x_1 \lor x_2 \lor x_3) \).
The first step in translating a 3-SAT problem to QUBO or Ising model is converting it to {CUBO} (Cubic Unconstrained Binary Optimization) or {CUSO} (Cubic Unconstrained Spin Optimization) formulations~\cite{Chancellor2016}, which are mathematically equivalent. 
CUSO translates Boolean SAT variables to spin variables \( s_i \in \{-1, +1\} \); CUBO, to binary variables \( x_i \in \{0, 1\} \).
Both formulations can describe cubic optimization problems such as 3-SAT, which feature at most third-order interactions among problem variables. 
There is a one-to-one correspondence between the SAT variables and the respective CUSO spins or CUBO variables. 
CUBO and CUSO formulations convert each SAT clause to a Hamiltonian, $H_B$ or $H_S$, respectively,
where $H_B= (H_S-7)/8$ applies. For the example clause \( (x_1 \lor x_2 \lor x_3) \):
\[
H_B = -x_1 - x_2 - x_3 + x_1x_2 + x_2x_3 + x_1x_3 - x_1x_2x_3
\]
\[
H_S = -s_1 - s_2 - s_3 + s_1s_2 + s_2s_3 + s_1s_3 - s_1s_2s_3
\]

\begin{table}[tb]
    \caption{
    Example 3-SAT clause $(x_1 \lor  x_2 \lor x_3)$ expressed in CUBO vs. CUSO format. 
    Hamiltonian energy ($H_B$ or $H_S$ respectively) reaches its maximum when the binary or spin variables fail to satisfy the clause.}
    \label{tab:truth_table_3SAT}
    \centering
     \begin{minipage}{0.35\linewidth}
     \small
        \centering
        \begin{tabular}{|c|c|c|>{\columncolor[gray]{1}}c|}
        \hline
        \multicolumn{4}{|c|}{\makecell{\textbf{CUBO} \\ \textbf{Cubic Unconstrained Binary Optimization}}} \\ \hline
        \rowcolor[gray]{0.9}
        $x_3$ & $x_2$ & $x_1$ & $H_B$ \\ \hline
        0  & 0  & 0  & 0 \\ \hline
        0  & 0  & 1  & -1 \\ \hline
        0  & 1  & 0  & -1 \\ \hline
        0  & 1  & 1  & -1 \\ \hline
        1  & 0  & 0  & -1 \\ \hline
        1  & 0  & 1  & -1 \\ \hline
        1  & 1  & 0  & -1 \\ \hline
        1  & 1  & 1  & -1 \\ \hline
        \end{tabular}
    \end{minipage}%
    \begin{minipage}{0.35\linewidth}
    \small
        \centering
        \begin{tabular}{|c|c|c|>{\columncolor[gray]{1}}c|}
        \hline
        \multicolumn{4}{|c|}{\makecell{\textbf{CUSO} \\ \textbf{Cubic Unconstrained Spin Optimization}}} \\ \hline
        \rowcolor[gray]{0.9}
        $s_3$ & $s_2$ & $s_1$ & $H_S$ \\ \hline
        -1  & -1  & -1  & +7 \\ \hline
        -1  & -1  & +1  & -1 \\ \hline
        -1  & +1  & -1  & -1 \\ \hline
        -1  & +1  & +1  & -1 \\ \hline
        +1  & -1  & -1  & -1 \\ \hline
        +1  & -1  & +1  & -1 \\ \hline
        +1  & +1  & -1  & -1 \\ \hline
        +1  & +1  & +1  & -1 \\ \hline
        \end{tabular}
    \end{minipage}
\end{table}
%
As illustrated in Table~\ref{tab:truth_table_3SAT},
the Hamiltonians $H_B$ or $H_S$ reach the maximum for the binary or spin assignments which fail to satisfy the clause (i.e., make it evaluate to logic one).
In other words, only assignments of binary or spin variables that satisfy the respective clause can minimize the clause-specific $H_B$ or $H_S$. 
This formulation gives rise to three different types of terms in the Hamiltonian equations: {\em diagonal} terms in the form of  \((x_1 \text{ or } s_1, \ldots)\); 
{\em quadratic} terms in the form of \((x_1x_2 \text{ or } s_1s_2, \ldots)\); and {\em cubic} terms in the form of \((x_1x_2x_3 \text{ or } s_1s_2s_3, \ldots)\). Quadratic and cubic terms represent quadratic (two-order or pairwise) and cubic (three-order) spin or variable interactions.
Ising machines rely on 
dedicated 
{couplers} to implement such spin or variable interactions in hardware.

\subsection{Diagonal, Quadratic, Cubic Coupler Logic}

\label{sec:Cubic Coupler}
Table \ref{fig:CUBO Solver} (Table \ref{fig:CUSO Solver}) provides the functional specification for 
a generic diagonal, quadratic, and 
cubic coupler considering CUBO (CUSO) formulation -- assuming positive coupling strengths, without loss of generality.
Each coupler's functional specification dictates the conditions for inter-variable coupling, systematically steering the OIM's evolution toward its global minimum.
By construction, these conditions result in the diagonal, quadratic or cubic terms that minimize a given Hamiltonian. 
The coupler logic 
must 
operate according to this 
functional specification. 
\begin{table}[ht]

\caption{Functional specification for diagonal, quadratic, and cubic couplers; assuming positive coupling strengths.}%
\label{fig:CouplerSpecs}

\centering
\setlength{\tabcolsep}{4pt}
\begin{subfigure}[t]{0.45\textwidth}
\small
\begin{minipage}[t]{0.45\textwidth}
\small
\raggedright
\textbf{Diagonal Coupler} \\[0.3em]
\begin{tabular}{|c|c|>{\centering\arraybackslash}p{2.2cm}|}
\hline
\textbf{$x_i$} & \textbf{Action} \\ \hline
0 & None \\ \hline
1 & Decouple $x_i$ \\ \hline
\end{tabular}

\vspace{0.5em}

\textbf{Quadratic Coupler} \\[0.3em]
\begin{tabular}{|c|c|>{\centering\arraybackslash}p{2cm}|}
\hline
\textbf{$x_i$} & \textbf{$x_j$} & \textbf{Action} \\ \hline
0 & 0 & None \\ \hline
0 & 1 & None \\ \hline
1 & 0 & None \\ \hline
1 & 1 & Decouple $x_i$ $x_j$ \\ \hline
\end{tabular}

\end{minipage}%
\begin{minipage}[t]{0.4\textwidth}
\small
\raggedright
\textbf{Cubic Coupler} \\[0.3em]
\begin{tabular}{|c|c|c|c|}
\hline
\textbf{$x_i$} & \textbf{$x_j$} & \textbf{$x_k$} & \textbf{Action} \\ \hline
0 & 0 & 0 & None \\ \hline
0 & 0 & 1 & None \\ \hline
0 & 1 & 0 & None \\ \hline
0 & 1 & 1 & None \\ \hline
1 & 0 & 0 & None \\ \hline
1 & 0 & 1 & None \\ \hline
1 & 1 & 0 & None \\ \hline
1 & 1 & 1 & Decouple $x_i$ $x_j$ $x_k$ \\ \hline
\end{tabular}
\end{minipage}

\vspace{.1cm}
\caption{{CUBO formulation.}}
\label{fig:CUBO Solver}

\end{subfigure}
\hfill
\begin{subfigure}[t]{0.5\textwidth}
\small

\begin{minipage}[t]{0.425\textwidth}
\small
\textbf{Diagonal Coupler} \\[0.3em]
\begin{tabular}{|c|c|>{\centering\arraybackslash}p{2.2cm}|}
\hline
\textbf{$s_i$} & \textbf{Action} \\ \hline
-1 & None \\ \hline
+1 & Decouple $s_i$ \\ \hline
\end{tabular}

\vspace{0.5em}

\textbf{Quadratic Coupler} \\[0.3em]
\begin{tabular}{|c|c|>{\centering\arraybackslash}p{2cm}|}
\hline
\textbf{$s_i$} & \textbf{$s_j$} & \textbf{Action} \\ \hline
-1 & -1 & Decouple $s_i$, $s_j$ \\ \hline
-1 & +1 & None \\ \hline
+1 & -1 & None \\ \hline
+1 & +1 & Decouple $s_i$ $s_j$ \\ \hline
\end{tabular} 
\end{minipage}%
\begin{minipage}[t]{0.6\textwidth}
\small
\raggedright
\textbf{Cubic Coupler} \\[0.4em]
\begin{tabular}{|c|c|c|c|}
\hline
\textbf{$s_i$} & \textbf{$s_j$} & \textbf{$s_k$} & \textbf{Action} \\ \hline
-1 & -1 & -1 & None \\ \hline
-1 & -1 & +1 & Decouple $s_i$,$s_j$,$s_k$ \\ \hline
-1 & +1 & -1 & Decouple $s_i$,$s_j$,$s_k$ \\ \hline
-1 & +1 & +1 & None \\ \hline
+1 & -1 & -1 & Decouple $s_i$,$s_j$,$s_k$ \\ \hline
+1 & -1 & +1 & None \\ \hline
+1 & +1 & -1 & None \\ \hline
+1 & +1 & +1 & Decouple $s_i$ $s_j$ $s_k$ \\ \hline
\end{tabular}
\end{minipage}

\vspace{.1cm}
\caption{{CUSO formulation.}}
\label{fig:CUSO Solver}
\end{subfigure}
\end{table}

{Diagonal couplers} 
establish coupling with a reference binary variable (Table \ref{fig:CUBO Solver}) or spin (Table \ref{fig:CUSO Solver}) -- which takes the form of a reference phase in OIM -- assuming a positive coupling coefficient and a reference state of 1 (Table \ref{fig:CUBO Solver}) or  +1 (Table \ref{fig:CUSO Solver}). Accordingly, the system 
favors states that correspond to 
0 (Table \ref{fig:CUBO Solver}) or \textminus 1 (Table \ref{fig:CUSO Solver})
aligning the variables or spins
with the expected minimum-energy configurations. 
%
Quadratic couplers
make the oscillators lock into the same phase ($s_i$ = $s_j$) for a positive coupling ($J_{ij}$ $>$ 0); and settle into opposite phases ($s_i$ = -$s_j$), for a negative coupling ($J_{ij}$ $<$ 0).
The end goal is 
minimizing 
the Ising Hamiltonian term $-J_{ij}$$s_i$$s_j$.
%
Tables \ref{fig:CUBO Solver} and \ref{fig:CUSO Solver} demonstrate this principle assuming positive coupling.

\begin{figure}
    \centering
    \includegraphics[width=0.37\linewidth]{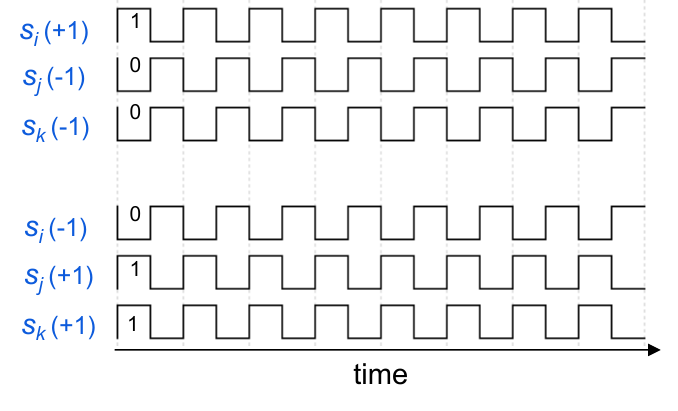}
    \caption{
    Spin configurations $s_is_js_k=-1+1+1$ 
and $s_is_js_k=+1-1-1$ 
from Table \ref{fig:CUSO Solver} produce indistinguishable oscillatory patterns.
Each spin maps to an oscillator, where the phase of the oscillator captures the spin state. Each square wave in the figure corresponds to a different oscillator, hence spin. A square wave of phase 0 encodes the spin state +1; of phase $\pi$, -1. For $s_is_js_k=-1+1+1$, the 3-spin 
system
oscillates between \(011\), \(100\), \(011\), ... over time; for $s_is_js_k=+1-1-1$, between \(100\), \(011\), \(100\), ... Hence, a pattern sample such as \(011\) observed at a given point in time may correspond to either spin configuration, $s_is_js_k=-1+1+1$ or $s_is_js_k=+1-1-1$.
}
    \label{fig:Waveform_cubic}
\end{figure}

The couplers can operate without a reference as long as they can distinguish oscillatory patterns corresponding to instantenous system state -- as tabulated in each row of Table \ref{fig:CUBO Solver} and Table \ref{fig:CUSO Solver} -- requiring different coupling actions from each other. This is not always the case for {quadratic} and {cubic} couplers: Figure \ref{fig:Waveform_cubic} demonstrates an example.
Spin configurations $s_is_js_k=-1+1+1$ 
and $s_is_js_k=+1-1-1$ 
produce indistinguishable oscillatory patterns, however, each configuration needs a different coupling action to force the system to a global minimum, per rows 3 and 6 of Table \ref{fig:CUSO Solver}, respectively.
This occurs because the Hamiltonian terms are symmetric under spin inversion.
As such oscillatory patterns encode spin states, the cubic coupler's only cue for identifying differences in spin configurations is differences in oscillatory patterns. 
To be more specific, for $s_is_js_k=-1+1+1$, the 3-spin 
system
oscillates between {\(011\), \(100\), \(011\), ... over time; for $s_is_js_k=+1-1-1$, between \(100\), \(011\), \(100\), ..., as Figure \ref{fig:Waveform_cubic} reveals. Hence, a pattern sample such as \(011\) observed at a given point in time may correspond to either spin configuration, $s_is_js_k=-1+1+1$ or $s_is_js_k=+1-1-1$.

In this case, identifying the absolute spin state requires a phase comparison against a reference, which dictates the control logic for coupling actions -- such as the control logic to enable or disable transmission gates, effectively modulating the inter-oscillator coupling necessary for system convergence.

In the context of OIMs, however, distributing the reference phase to each (tranmission gate based) coupler across a densely packed oscillator network is challenging.

This ambiguity applies to quadratic CUBO couplers (Table \ref{fig:CUBO Solver}), as well, but not to quadratic CUSO couplers (Table \ref{fig:CUSO Solver}):
Configurations $x_i x_j = 00$ and $x_i x_j = 11$ from Table \ref{fig:CUBO Solver}
produce indistinguishable oscillatory patterns.
However, while $x_i x_j = 00$ does not require any action, $x_i x_j = 11$ calls for decoupling the respective oscillator phases.
%
Without a reference,
the coupler cannot identify which state the system is in (i.e., which $x_i x_j$ configuration applies) and take the corresponding action accordingly. 
Addressing this requires 
comparison 
to a known reference. For example, if the {\em reference} is 1, the decoupling condition for $x_i x_j = 11$ can be expressed as 
\((x_i \, \text{XNOR} \, \text{\em reference}) \, \text{AND} \, (x_j \, \text{XNOR} \, \text{\em reference})\) -- to control, e.g., transmission gate activation accordingly.
This discrepancy does not apply to the quadratic coupler from Table \ref{fig:CUSO Solver}, because the coupling actions for $s_i s_j = -1 -1$ and  $s_i s_j = 11$ are identical. 
}


\subsection{Higher Order (Beyond Cubic) Coupling Logic}
\label{sec:Quartic Coupler}



\begin{table}[h]
\centering
\caption{Functional specification for the proposed Quartic coupler, which eliminates any ambiguity in oscillatory patterns that necessitates a resolution with a known reference. }
\label{tab:Quartic_Unit}
\small
\begin{tabular}{ccccccc}
\toprule
\textbf{$s_i$} & \textbf{$s_j$} & \textbf{$s_k$} & \textbf{$s_l$} & \textbf{Binary Mapping} & \textbf{Parity} & \textbf{Action} \\ \midrule
\textbf{-1} & \textbf{-1} & \textbf{-1} & \textbf{-1} & \textbf{0000} & \textbf{0} & \textbf{Decouple $s_1 s_2 s_3 s_4$} \\
-1 & -1 & -1 & +1 & 0001 & 1 & None \\
-1 & -1 & +1 & -1 & 0010 & 1 & None \\
\textbf{-1} & \textbf{-1} & \textbf{+1} & \textbf{+1} & \textbf{0011} & \textbf{0} & \textbf{Decouple $s_1 s_2 s_3 s_4$} \\ \midrule
-1 & +1 & -1 & -1 & 0100 & 1 & None \\
\textbf{-1} & \textbf{+1} & \textbf{-1} & \textbf{+1} & \textbf{0101} & \textbf{0} & \textbf{Decouple $s_1 s_2 s_3 s_4$} \\
\textbf{-1} & \textbf{+1} & \textbf{+1} & \textbf{-1} & \textbf{0110} & \textbf{0} & \textbf{Decouple $s_1 s_2 s_3 s_4$} \\
-1 & +1 & +1 & +1 & 0111 & 1 & None \\ \midrule
+1 & -1 & -1 & -1 & 1000 & 1 & None \\
\textbf{+1} & \textbf{-1} & \textbf{-1} & \textbf{+1} & \textbf{1001} & \textbf{0} & \textbf{Decouple $s_1 s_2 s_3 s_4$} \\
\textbf{+1} & \textbf{-1} & \textbf{+1} & \textbf{-1} & \textbf{1010} & \textbf{0} & \textbf{Decouple $s_1 s_2 s_3 s_4$} \\
+1 & -1 & +1 & +1 & 1011 & 1 & None \\ \midrule
\textbf{+1} & \textbf{+1} & \textbf{-1} & \textbf{-1} & \textbf{1100} & \textbf{0} & \textbf{Decouple $s_1 s_2 s_3 s_4$} \\
+1 & +1 & -1 & +1 & 1101 & 1 & None \\
+1 & +1 & +1 & -1 & 1110 & 1 & None \\
\textbf{+1} & \textbf{+1} & \textbf{+1} & \textbf{+1} & \textbf{1111} & \textbf{0} & \textbf{Decouple $s_1 s_2 s_3 s_4$} \\ \bottomrule
\end{tabular}
\end{table}

It is possible to support higher order interactions without compromising hardware complexity: Table \ref{tab:Quartic_Unit} reveals 
the proposed {quartic} coupler design which can represent (up-to) fourth-order interactions and which does not need a known reference for correct operation.
Specifically, the quartic coupler can distinguish configurations that require different coupling actions from each other by solely checking the parity of the configuration. As shown in Table \ref{tab:Quartic_Unit}, each of the two possible parity values is uniquely associated with a specific coupling action. Hence, there is no need for a reference.
The quartic 
coupler only needs to decouple spins when their combined parity is 0. No (de)coupling action is necessary otherwise.

In principle, for each COP, the OIM gets programmed with the corresponding coupling coefficients $J$  
once at the beginning of the optimization. Coupling coefficients remain constant throughout execution. A dedicated quartic coupler controls each group of four spins, and periodically evaluates the parity of the respective four spins according to Table \ref{tab:Quartic_Unit}. Only if the parity is 0, the coupler ``nudges'' -- i.e., temporarily decouples -- the four spins to stimulate a state change. 
Spins continuously evolve as the coupled spins relax under the collective influence of all active couplers. 

The quartic coupler can thereby enable practical Higher Order Oscillatory Ising Machines (\arch) which can seamlessly map COPs featuring higher order interactions.

The core idea behind the parity-based quartic coupler design naturally extends to any even number of interactions beyond four.
At the same time, the quartic coupler can realize all lower order couplings than fourth-order:
The idea is using an extra (termed dummy) variable to translate a lower-order clause to a higher-order one. 
For example, we can translate
the 3-SAT clause \( (x_1 \lor x_2 \lor x_3) \) 
into 4-SAT by introducing a dummy variable $d$:
\( (x_1 \lor x_2 \lor x_3 \lor d) \) $\land$ \( (x_1 \lor x_2 \lor x_3 \lor \lnot d) \).
$d$ does not influence the logical outcome of the original clause -- 
regardless of the truth value of \( d \), 
\( (x_1 \lor x_2 \lor x_3  \lor d) \) $\land$ \( (x_1 \lor x_2 \lor x_3 \lor  \lnot d) \) always
preserves the satisfiability of the original 3-SAT clause \( (x_1 \lor x_2 \lor x_3) \). 



\subsection{Practical Considerations}
\label{sec:Practical Considerations}

In structured and sparse topologies such as the King's graph, for both pairwise and quartic designs, the number of couplers scale linearly with the number of spins $N$ (which represents a good proxy for problem size), i.e., $\mathcal{O}(N)$.
By definition, however, quartic couplers reduce the total number of (physical) spins necessary to map a given problem, which in turn leads to a {net reduction in silicon area} devoted to oscillators.
Consequently, parity-based higher-order couplers are ideally suited for topologies such as the King’s graph, where the coupler-to-spin ratio scales sub-linearly as the problem size increases. 

\begin{figure}[htp]
\centering

\begin{minipage}[t]{0.5\linewidth}
\vspace{0pt}
\centering

\includegraphics[
width=0.85\linewidth,
height=7.5cm,
keepaspectratio
]{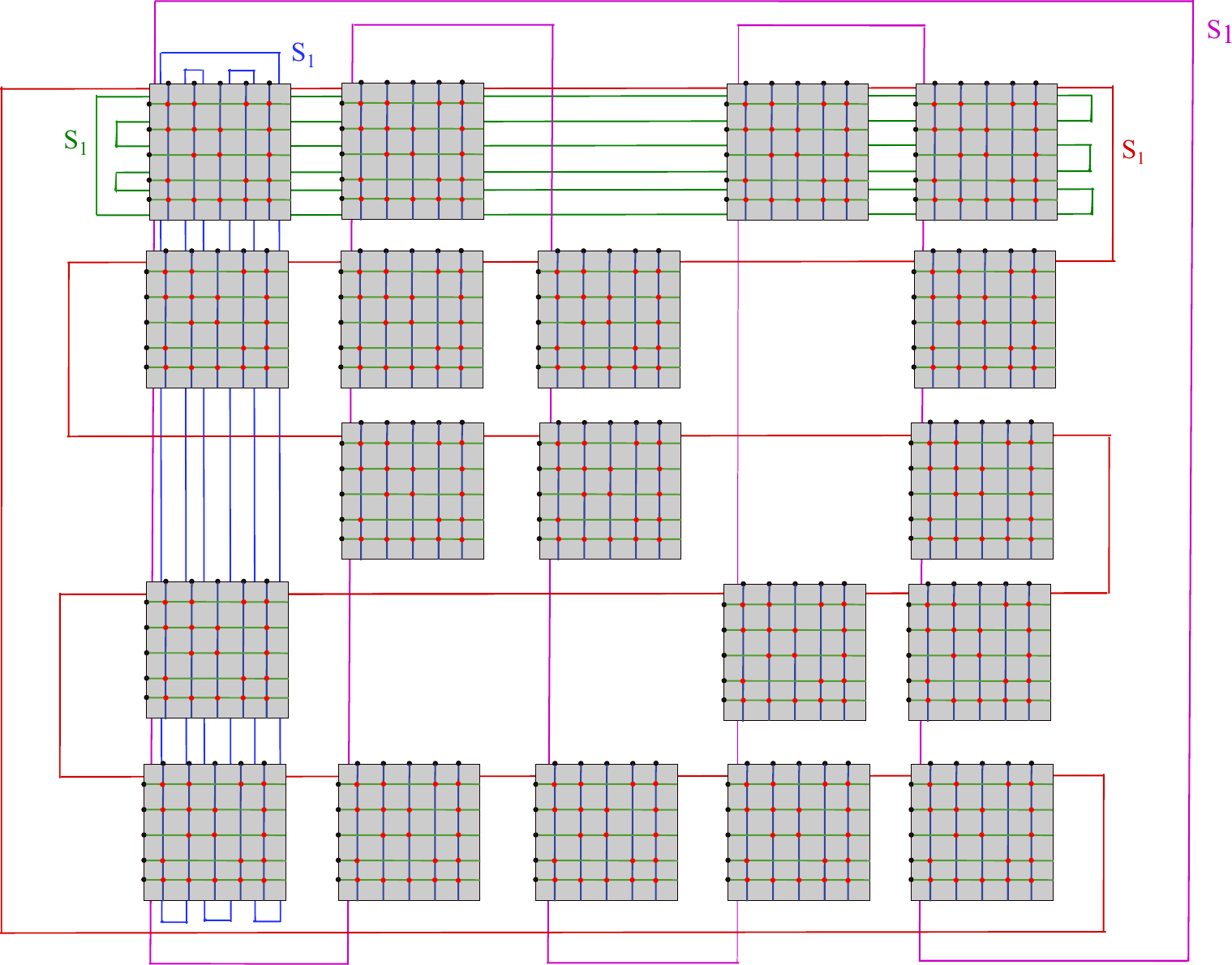}

\vspace{2mm}
(a)
\end{minipage}
\begin{minipage}[t]{0.45\linewidth}
\vspace{5mm}
\centering
\includegraphics[
width=2.5cm,
keepaspectratio
]{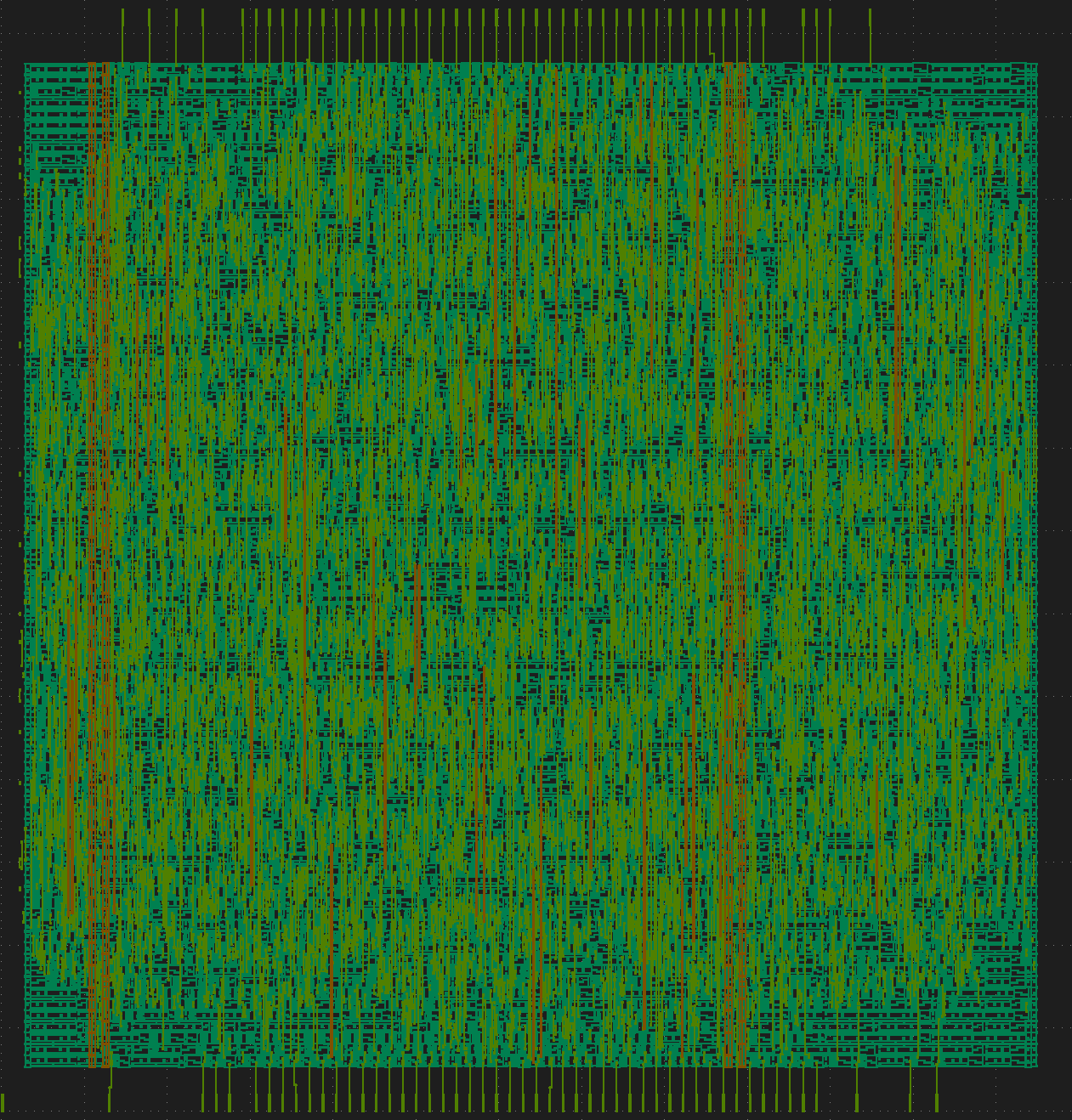}

\vspace{1mm}
(b)

\vspace{5mm}

\includegraphics[
width=1.4cm,
keepaspectratio
]{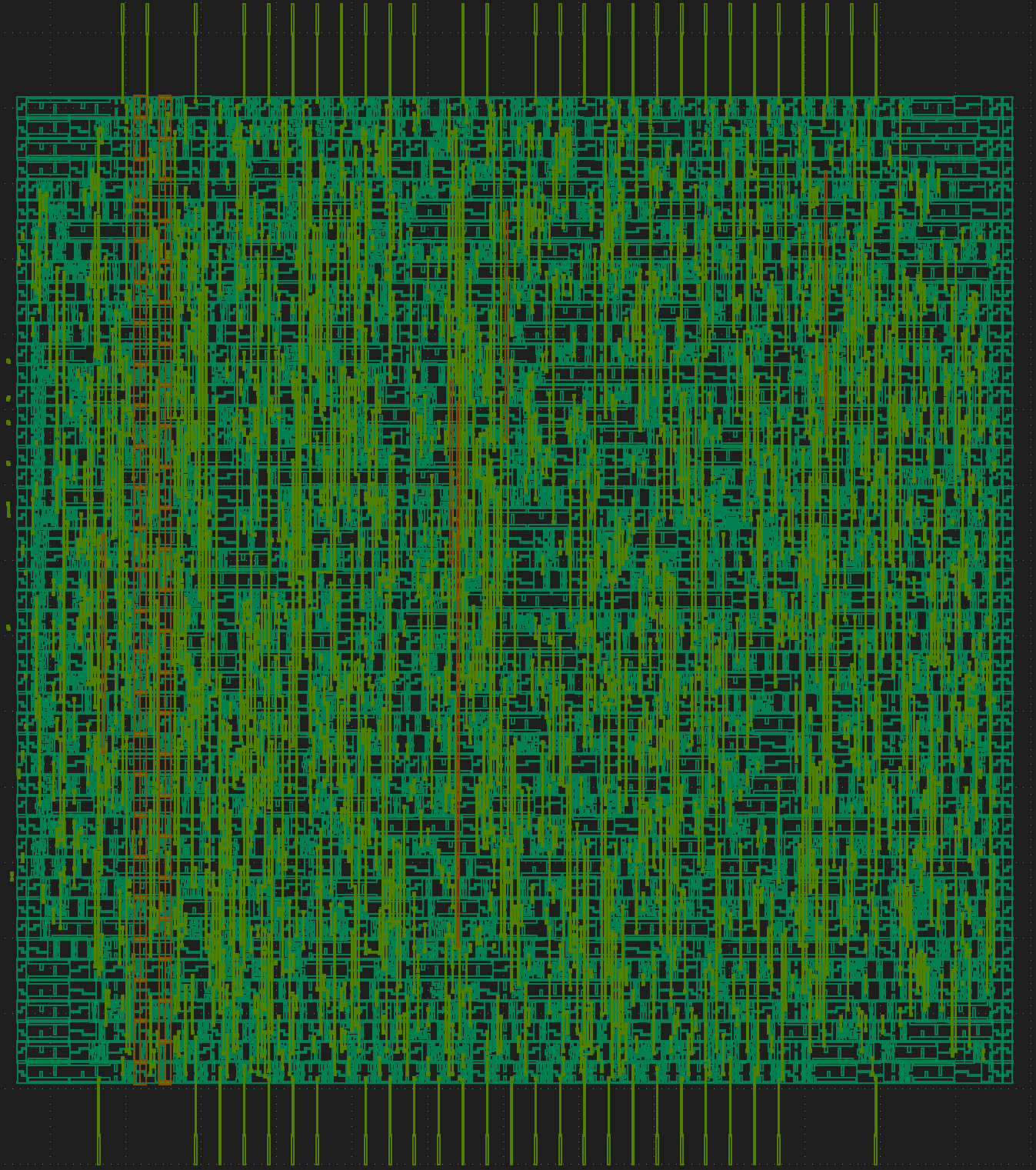}

\vspace{1mm}
(c)

\end{minipage}

\caption{(a) Quartic coupling on King's graph. Post-routing layouts of (a) a representative reference-assisted POIM vs. (b) the proposed reference-free HOIM architecture.
Layouts are at approximately the same scale: HOIM occupies $67.6\%$ less die area than POIM.}  

\label{fig:Hardware_and_layout}
\end{figure}

Figure~\ref{fig:Hardware_and_layout} captures an illustrative example based on the 
representative POIM design from~\cite{lo2023ising}, where
spins are realized as ring oscillators (a few inverter stages) connected in a crossbar.
Positive and negative couplings are enforced by connecting oscillators in-phase or out-of-phase, respectively. Each spin has horizontal and vertical replicas, and every crossbar intersection hosts a transistor-based coupler whose control signal determines whether spins $s_i$ and $s_j$ are coupled.
Extending this design to fourth-order coupling requires four replicas per spin (along horizontal, vertical, and two diagonal directions), as demonstrated by Figure~\ref{fig:Hardware_and_layout}a for a 5-spin excerpt to ease illustration.
Diagonal spins (NE–SW and NW–SE) traverse all couplers within each block (green and blue lines), while horizontal and vertical spins follow the pink and red paths. 
The unique intersection of all four replicas defines a quartic coupling cell. 
In contrast to POIM, where pairwise couplers reside at every spin 
interaction point of an $N \times N$ crossbar (red dots), the quartic 
implementation requires $DN$ such blocks, where $D$ denotes the average 
degree of the underlying hardware topology. Since each spin 
participates in $D$ quartic interactions, and each interaction must be 
evaluated by a dedicated block to feed back its contribution to the 
spin dynamics, the per-spin hardware overhead scales as $D$ coupling 
blocks, resulting in a total of $DN$ blocks across the architecture.

\section{Evaluation Setup} 
\label{sec:Setup}

\noindent \textbf{Simulation Framework:} 
We simulate POIM and HOIM
using
a QUBO/HUBO solver based on the
the Dual Annealing optimization algorithm~\cite{scipy2024dualannealing}, which
integrates the general principles of Classical Simulated Annealing (CSA) and Fast Simulated Annealing (FSA), enhanced with effective local minima escape strategies. 
For our experiments, we convert 3-SAT instances to QUBO using the Chancellor's formulation~\cite{Chancellor2016}.
We derive the corresponding HUBO formulae using the principles
outlined 
in 
Section \ref{sec:Implementation}.
%
%
Our simulation framework supports all-to-all (A2A) connected OIMs~\cite{lo2023ising,9204635}, as well as 
OIMs with nearest neighbor connectivity~\cite{9377463,7350099}.
To ensure hardware compatibility, we normalize each Hamiltonian to satisfy the constraints of the simulated OIM. 
We model interaction coefficients after a representative POIM~\cite{lo2023ising}. Each iteration on this baseline POIM takes approximately 171.9$\mu$s and consumes 10mW of power.
Since the simulator does not account for the stochastic noise and hardware non-idealities present in physical implementations, the reported results establish an upper bound on achievable performance.



{

\noindent \textbf{Benchmarks:} 
We use benchmark instances from the SATLIB suite~\cite{SATLIB}: uf20-91 (20 variables/91 clauses) and uf50-218 (50 variables/218 clauses). All instances considered are satisfiable and
come from the phase transition region~\cite{10.1007/3-540-45578-7_11} where the hardest SAT problems reside. 
We also use randomly generated uniform 4-SAT instances of 20 and 50 variables
from the phase transition region --  where each problem is satisfiable and each possible literal is selected with the same probability of \(1/2n \); $n$ being the number of problem variables. 

\section{Evaluation} 
\label{sec:Results}
\noindent {\textbf{Performance Impact:}}
{
We start our analysis 
with the HOIM vs. POIM comparison of the average iteration count required to reach the All-SAT condition (where all clauses are satisfied)
considering problem instances from SATLIB 3-SAT uf20-91 in Figure~\ref{fig:AverageIter20}; and uf50-218, in Figure~\ref{fig:AverageIter50}.}


\begin{figure}[htp]
    \centering   
    \begin{subfigure}{0.49\textwidth}
        \centering
        \includegraphics[width=\linewidth]{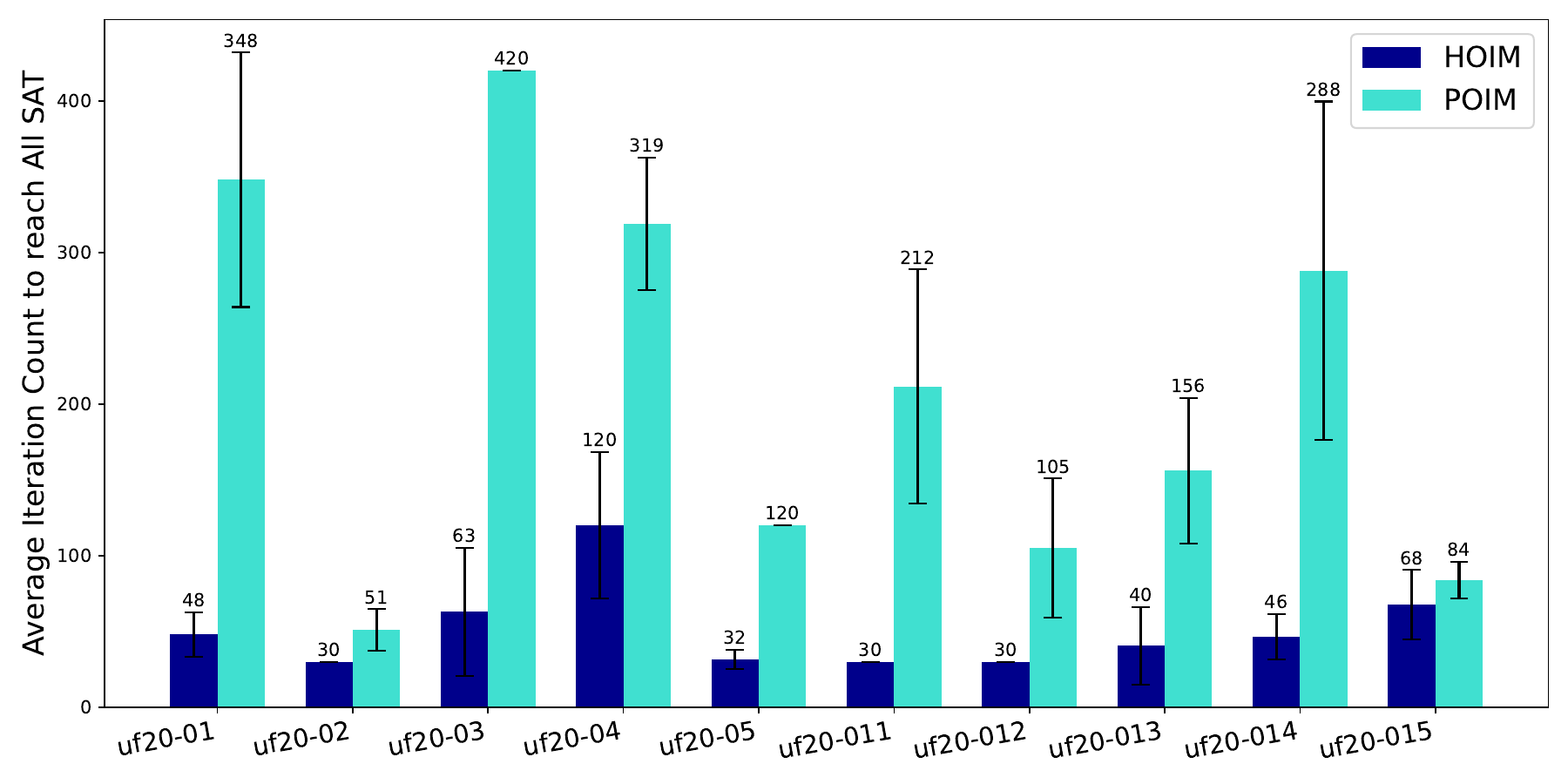}
        \caption{uf20-91}
        \label{fig:AverageIter20}
    \end{subfigure}
    \hfill
    \begin{subfigure}{0.49\textwidth}
        \centering
        \includegraphics[width=\linewidth]{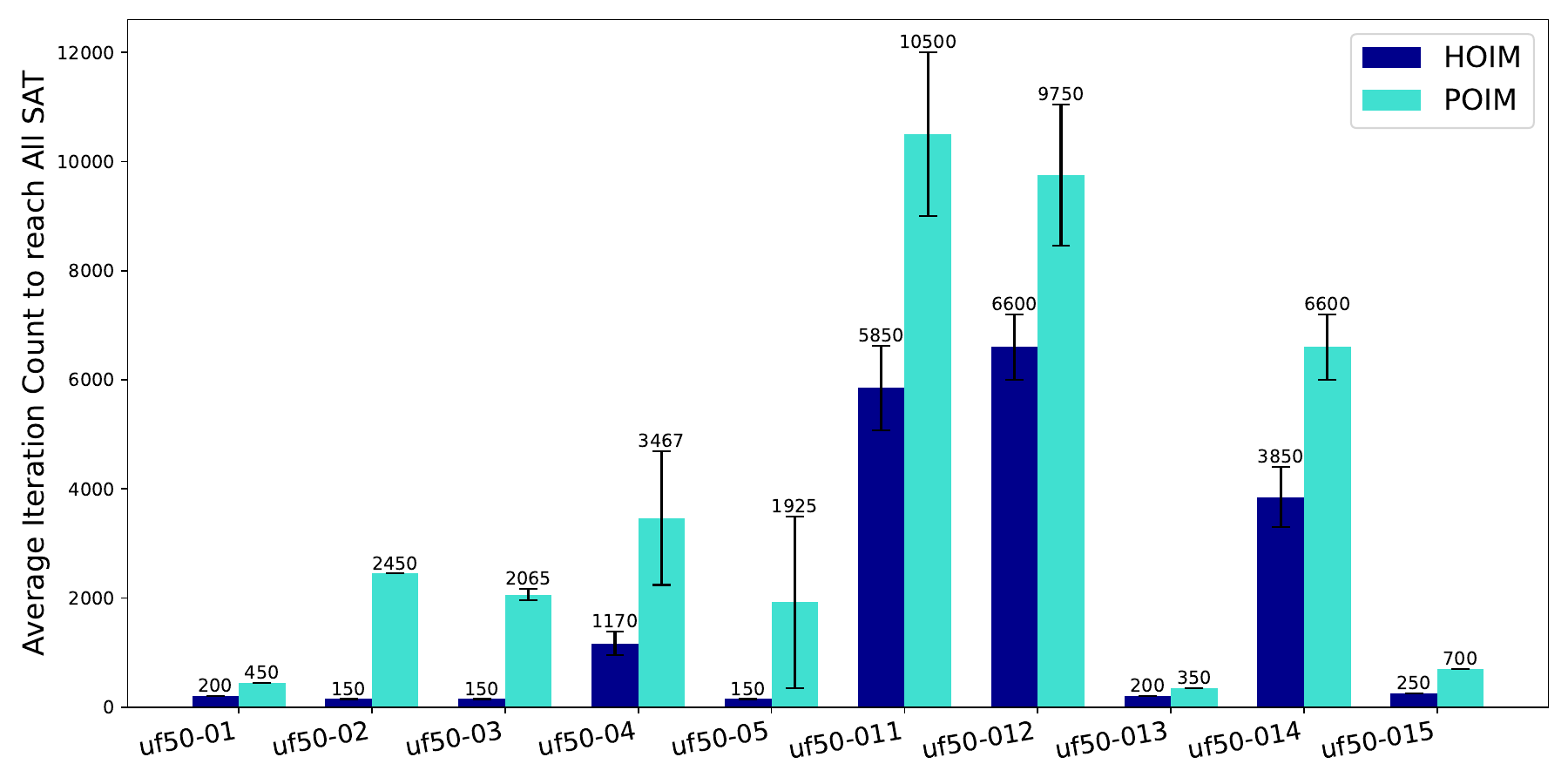}
        \caption{uf50-218}
        \label{fig:AverageIter50}
    \end{subfigure}
    
    \caption{Average number of iterations to reach All-SAT.}
    \label{fig:AverageIter}
\end{figure}


{We observe that HOIM consistently outperforms POIM by requiring significantly fewer iterations in most cases. For example, for uf20-01 and uf20-03, POIM requires approximately 350–420 iterations, while HOIM only needs about 50–65. The only notable exception is uf20-015. Overall, HOIM requires about 4.3$\times$ fewer iterations on average compared to POIM. A similar trend applies to uf50-218, where HOIM requires ~6$\times$ fewer iterations than POIM. Additionally, in both cases HOIM shows lower variance, 
suggesting more stable performance.}

\begin{figure}[t]
\centering
\begin{subfigure}[t]{0.3\textwidth}
\centering
\vspace{0pt}
\includegraphics[width=.9\linewidth]{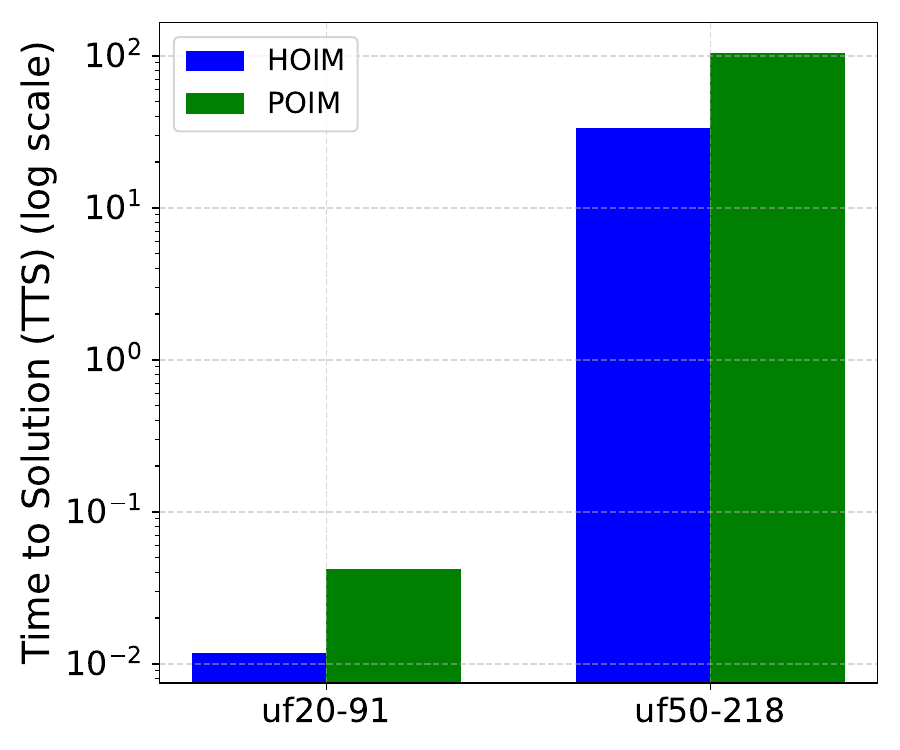}
\caption{TTS for 3-SAT problems.}
\label{fig:Metric_SATLIB_TTS}
\end{subfigure}
\hfill
\begin{subfigure}[t]{0.3\textwidth}
\centering
\vspace{0pt}
\includegraphics[width=.9\linewidth]{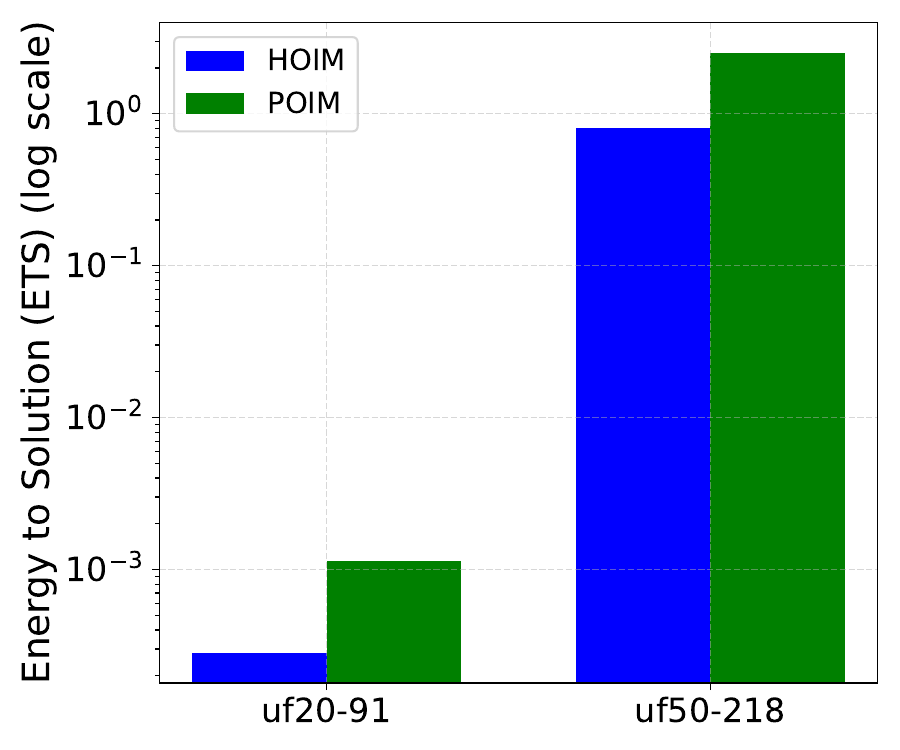}
\caption{ETS for 3-SAT problems.}
\label{fig:Metric_SATLIB_ETS}
\end{subfigure}
\hfill
\begin{subfigure}[t]{0.3\textwidth}
\centering
\vspace{0pt}
\includegraphics[width=.9\linewidth]{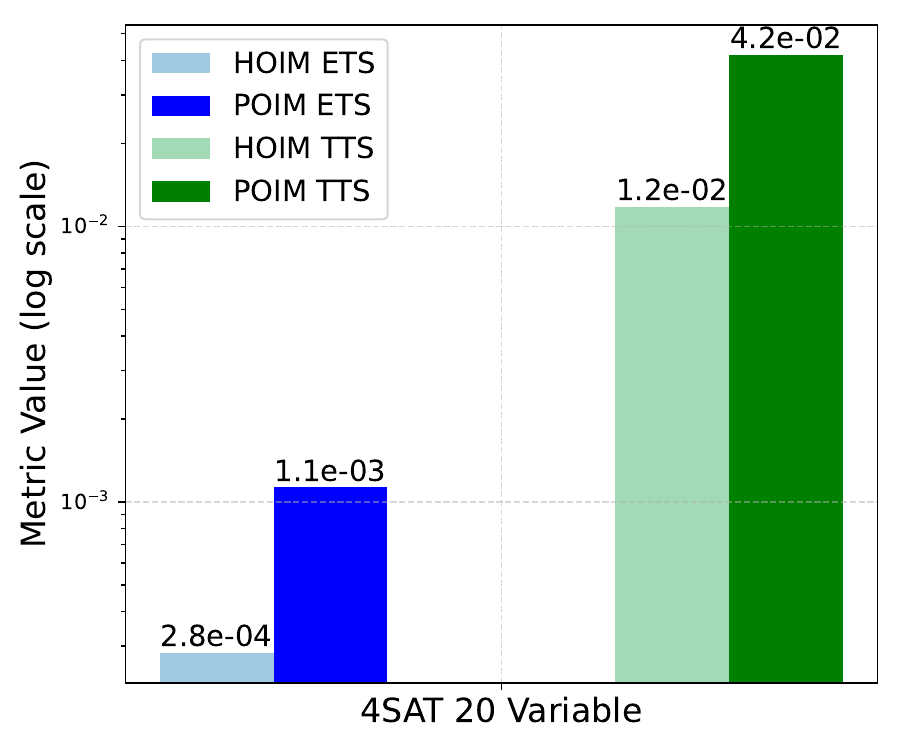}
\caption{TTS and ETS for 4-SAT problems.}
\label{fig:Metrices}
\end{subfigure}
\caption{{Time and Energy To Solution.}}
\label{fig:PerformanceMetrics}
\end{figure}

Figure~\ref{fig:Metric_SATLIB_TTS} and~\ref{fig:Metric_SATLIB_ETS} provide a direct performance comparison of HOIM to POIM in terms of Time to Solution (TTS) and Energy to Solution (ETS) for 20- and 50-variable SATLIB problems. Time to Solution (TTS)~\cite{Pelofske2022ParallelQA} quantifies the expected time required for a solver to obtain the correct (ground-state) solution at least once with a specified confidence level. Energy to Solution (ETS) measures the total energy consumption required to reach a correct or acceptable solution with a target confidence level. We use a confidence level of $90\%$.
We observe that enabling higher-order coupling (HOIM) reduces TTS and ETS by up-to 4$\times$ in 20-variable 3-SAT problems; and upto ~3$\times$ in 50-variable 3-SAT problems. 
As depicted by Figure~\ref{fig:Metrices}, for 4-SAT, the reduction in TTS and ETS reaches by up-to 7.89$\times$.



\begin{figure} [htp]
    \centering
    \includegraphics[width=0.33\columnwidth]{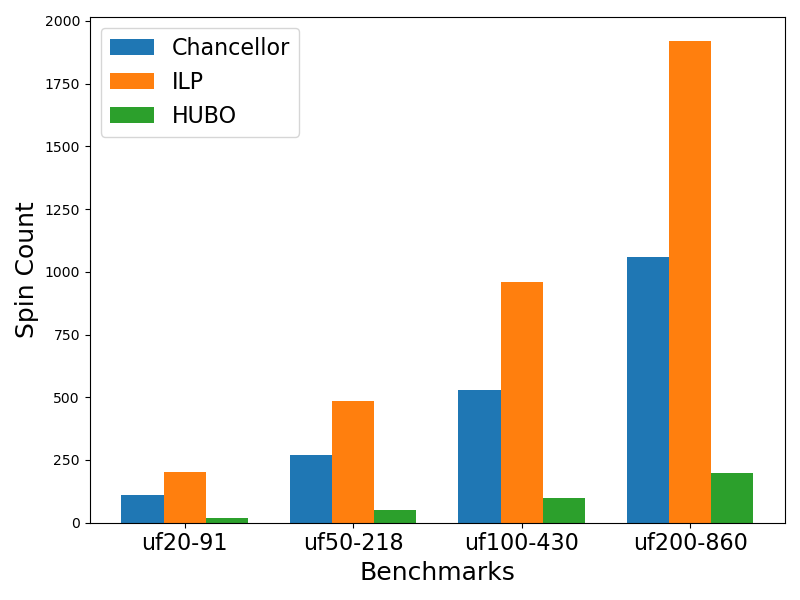}
    \caption{
    Ising spin (equivalently, QUBO variable) count in the translated problem as a function of problem size, using different formulations for mapping higher-order SAT interactions.
x-axis covers 3-SAT problems from SATLIB~\cite{SATLIB} of increasing problem size.}
    \label{fig:spincount}
\end{figure}

\noindent {\textbf{Impact on Physical Spin Count \& Decomposition:}
OIM spin count cannot increase indefinitely due to fundamental physical limits.
However, COP sizes of practical importance keep growing. 
Ancillary variables introduced in the process of converting higher-order problem interactions to lower-order (pairwise) machine interactions exacerbate this by shrinking the effective number of available hardware spins.
Large problems that cannot fit into a given Ising machine must be {\em decomposed}, i.e., broken down into smaller subproblems that match the available spin count in hardware. The complication comes from such subproblems not being independent. As a result, decomposition can degrade solution accuracy and/or increase the time to reach a solution.   
The subproblems are solved iteratively until 
either all clauses are satisfied (reaching the {All-SAT} condition) 
or when a predefined iteration limit is reached.
HOIM, by construction, reduces the pressure on decomposition and its adverse impact on solver performance.

\begin{figure}[tph]
\centering

\begin{subfigure}[t]{0.35\textwidth}
\centering
\includegraphics[width=.81\linewidth]{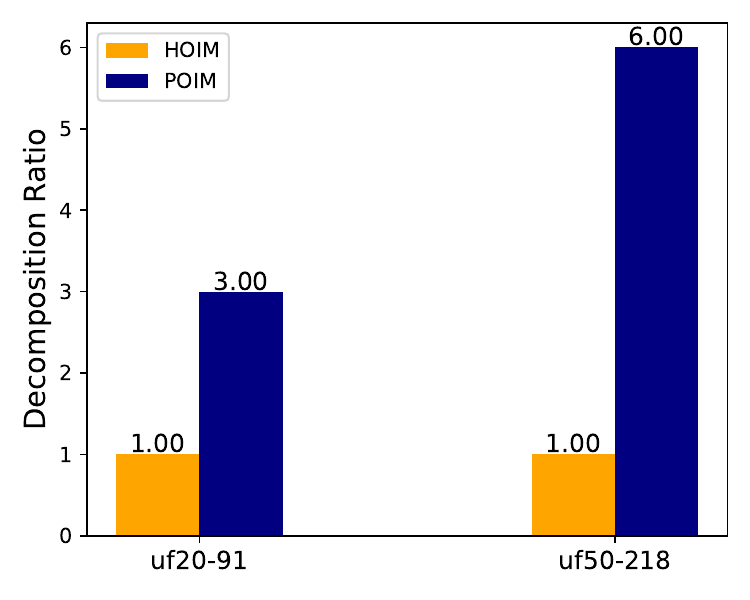}
\caption{Decomposition ratio}
\label{fig:Decomp_ratio}
\end{subfigure}
\hspace{0.04\textwidth}
\begin{subfigure}[t]{0.35\textwidth}
\centering
\includegraphics[width=.81\linewidth]{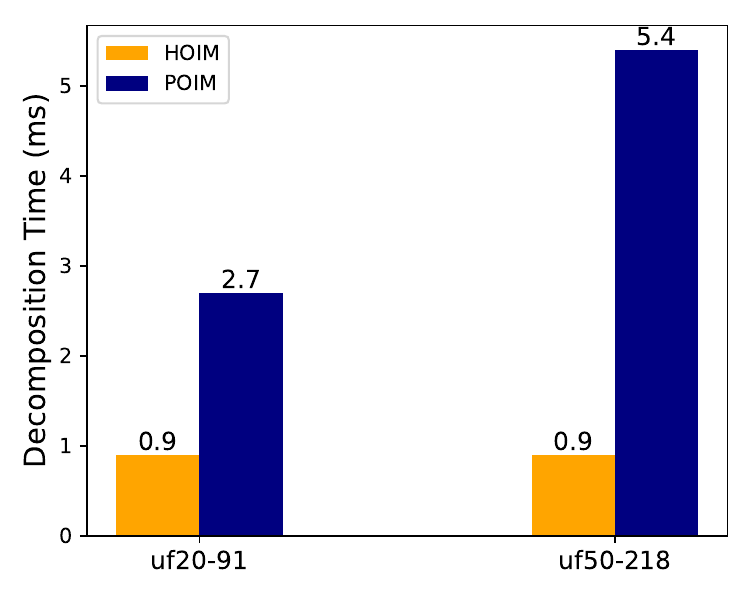}
\caption{Decomposition time}
\label{fig:Decomp_time}
\end{subfigure}

\caption{{Impact on decomposition.}}
\label{fig:Decomposition}

\end{figure}

Figure \ref{fig:spincount} shows  
spin count in the translated problem as a function of problem size, using different formulations for mapping higher-order SAT interactions. 
The x-axis covers 3-SAT problems from SATLIB~\cite{SATLIB} of increasing problem size -- 20, 50, 100, and 200 variables.
We consider QUBO (Chancellor's and ILP) and HUBO formulations.
{The plot clearly shows that as the problem size increases from 20 to 200 variables, the HUBO formulation significantly reduces the overall spin count by eliminating the need for ancillary spins. 
Specifically, HUBO requires $\approx 5.5\times$ fewer spins than the most efficient QUBO formulation ({Chancellor's}) for 20 variable problems and $\approx 5.3\times$ for 200 variable problems.

Furthermore, due to hardware constraints~\cite{lo2023ising}, even a 20-variable 3-SAT problem cannot be directly mapped to the OIM chip after translation to the QUBO format, necessitating decomposition into smaller sub-problems. In contrast, formulating the problem using higher-order interactions (HUBO) avoids the introduction of ancillary variables, substantially reducing the need for decomposition by 3$\times$(20 variables) to 6$\times$ (50 variables), as captured by the decomposition ratio in Figure \ref{fig:Decomp_ratio}. 
{The decomposition ratio quantifies the number of times a QUBO formulation must be decomposed compared to the corresponding HUBO. For example, if a HUBO problem can be solved on hardware 
without decomposition, while the equivalent QUBO problem requires splitting into a minimum of three subproblems, the decomposition ratio becomes 3$\times$. The increasing decomposition ratio also affects the time spent in decomposition as shown in Figure~\ref{fig:Decomp_time}.
}

\begin{figure}[thp]
\centering
\begin{subfigure}{0.49\textwidth}
    \centering
    \includegraphics[width=\linewidth]{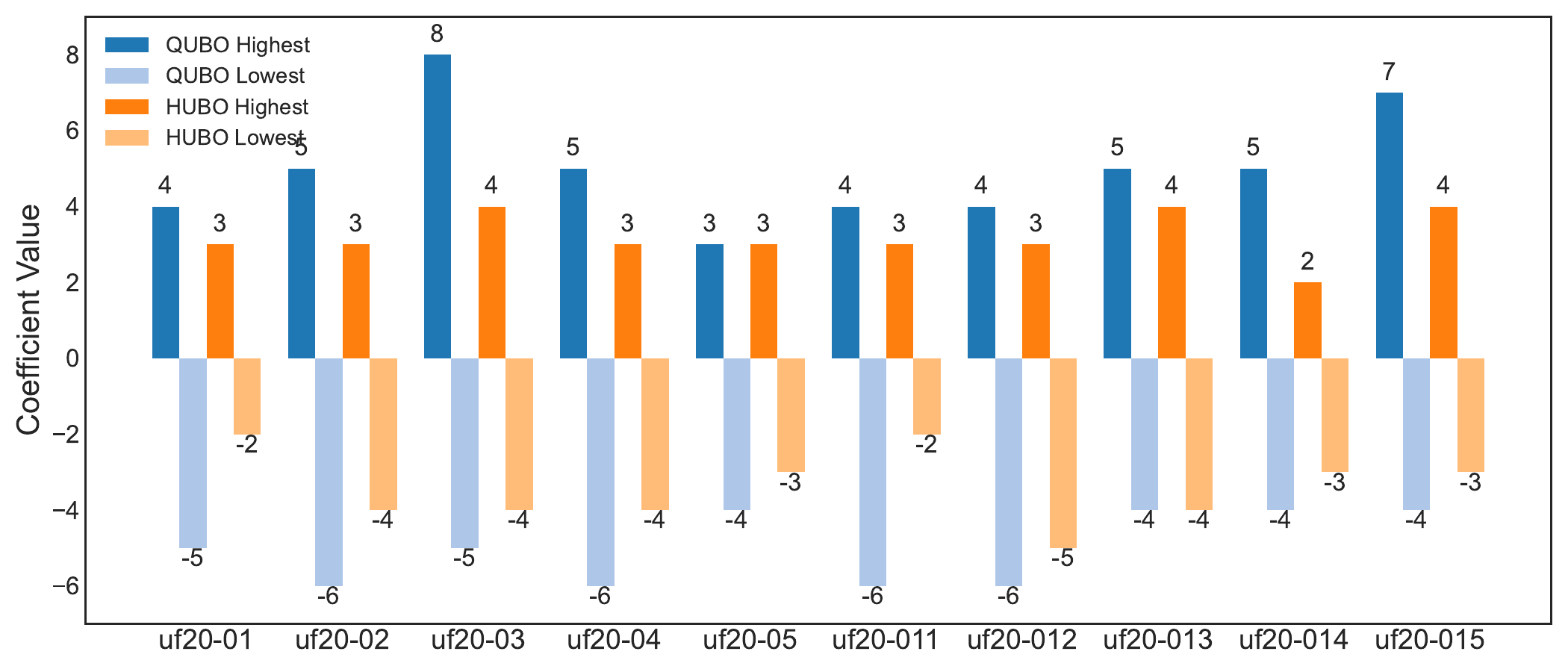}
    \caption{uf20-91 instances}
    \label{fig:Coeff_20variable}
\end{subfigure}
\hfill
\begin{subfigure}{0.49\textwidth}
    \centering
    \includegraphics[width=\linewidth]{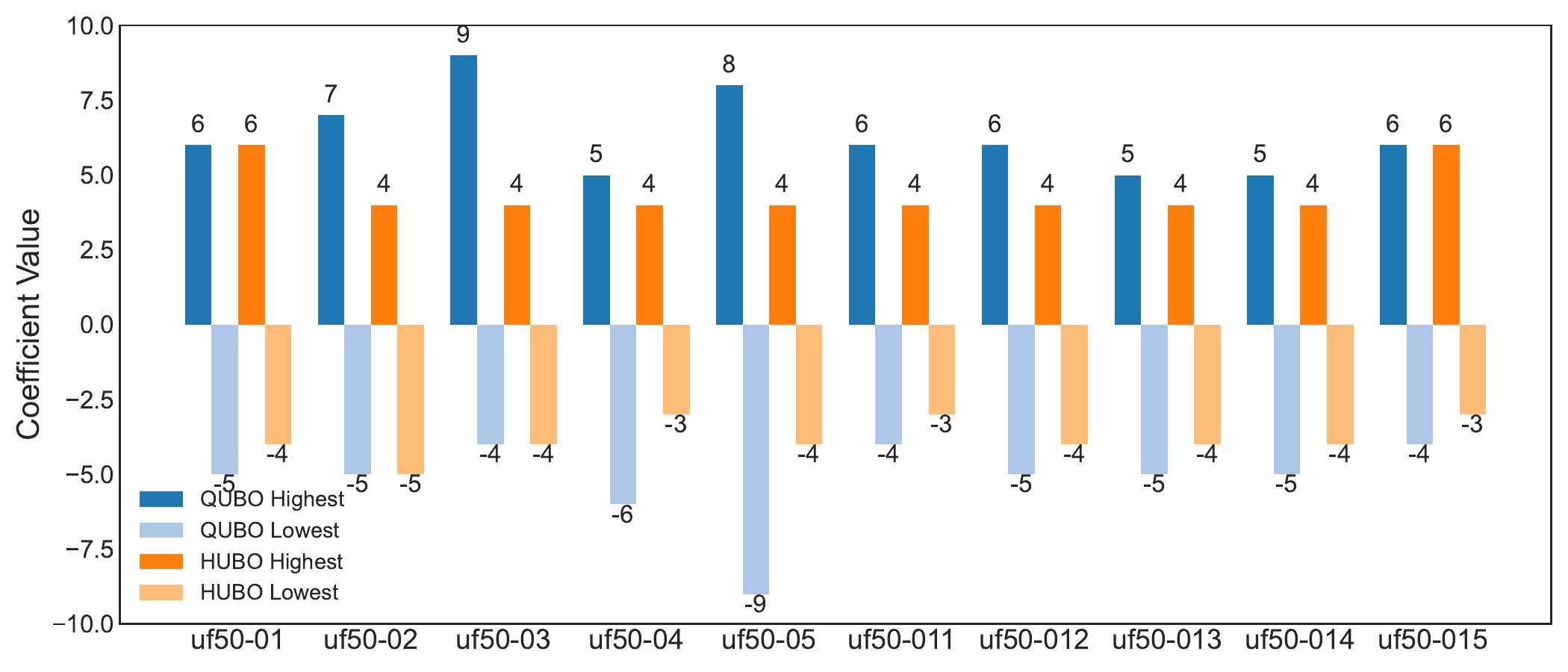}
    \caption{uf50-218 instances}
    \label{fig:Coeff_50variable}
\end{subfigure}
\vspace{1em}
\begin{subfigure}{0.7\textwidth}
    \centering
    \includegraphics[width=.5\textwidth]{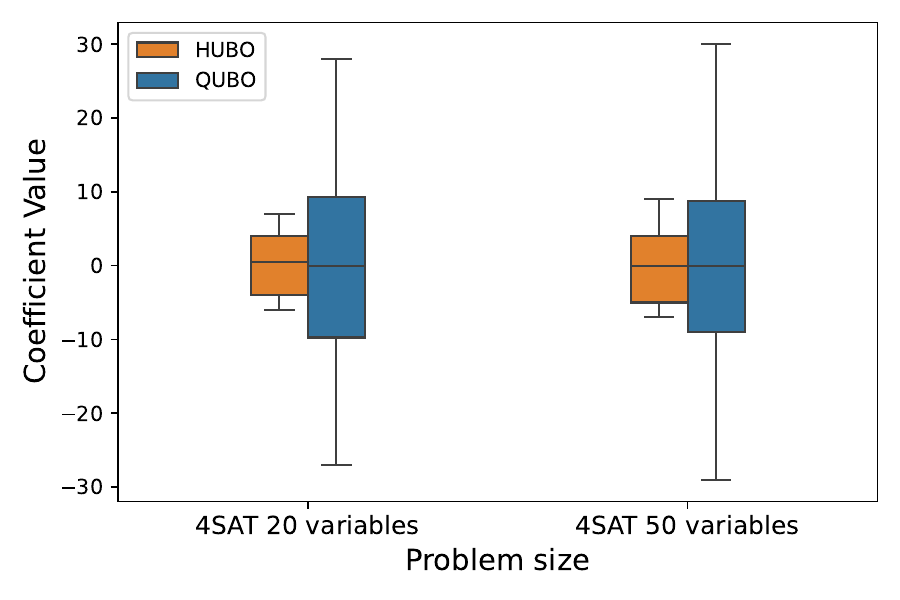}
    \caption{20 and 50 variable random 4-SAT problems}
    \label{fig:Coeff_dist}
\end{subfigure}
\caption{Coefficient ranges for different SAT benchmarks.}
\label{fig:CoeffComparison_All}
\end{figure}

\noindent {\textbf{Impact on Coefficient Range:}}
{
Typical OIMs are not just limited by the number of spins. The range and precision for interaction coefficients are also limited. 
The process of converting higher-order problem interactions to lower-order (pairwise) machine interactions boils down to another layer of problem translation and typically affects the coefficient range as well as the precision. This may result in coefficients smaller or larger than the underlying machine can support, necessitating
excessive scaling and/or rounding of coefficient values, which in turn may degrade the solution accuracy.
By simplifying problem translation, \arch\ can reduce the likelihood of excessive scaling {and/or} rounding, as well.
Figure~\ref{fig:Coeff_20variable} captures this effect for 20-variable 3-SAT problems; Figure~\ref{fig:Coeff_50variable}, for 50-variable 3-SAT problems; and Figure~\ref{fig:Coeff_dist}, for 4-SAT problems, respectively.

%
We observe that 
the QUBO formulation (Chancellor~\cite{Chancellor2016}) exhibits a broader coefficient range compared to HUBO for 20- and 50-variable problems. For example, QUBO reaches highs of 8, 7, and 5 in 20-variable problems, while HUBO's highest values are mostly around 4 or 3. As problem size increases to 50 variables, the same trend persists. Even for denser problems like 4-SAT where clause to variable ratio is 2$\times$ greater than 3-SAT, QUBO range is significantly higher, with coefficients stretching from about –30 to 30 for 50-variable problems, compared to HUBO's more restrained range of –10 to 9. Even in smaller 20-variable 4-SAT instances, QUBO coefficients span  –30 to 28, while HUBO coefficients remain between –8 and 10.  
We conclude that QUBO 
is more likely to hit hardware limits for interaction coefficient range and precision than HUBO.



\noindent {\textbf{Impact on Chip Area:}}
{We implement and physically synthesize two inverter-ring-oscillator-based architectures using the Sky130 CMOS process and the OpenLane ASIC flow. The first architecture is a representative reference-assisted OIM consisting of 50 spins, augmented to support quartic coupling by introducing two additional spin signal lines (the diagonal and anti-diagonal ring families) routed to each coupler. The second architecture is the proposed reference-free HOIM consisting of 50 spins, supporting quartic coupling without per-coupler reference-assisted phase comparison. Both architectures have King's graph topology.}

%
Both designs are synthesized under identical conditions at 45\% core utilization. The proposed reference-free HOIM achieves a 67.6\% reduction in die area and a 70.8\% reduction in core area relative to the reference-assisted baseline. Routing complexity is reduced even more substantially: total routed wire-length is reduced by 74.4\%, total routed nets by 75.4\%, and total vias by 75.6\%, with the maximum single-net wire-length reduced by 38.6\%.

The reference-assisted architecture requires dedicated routing paths for phase comparison across the oscillator network, increasing interconnect complexity and routing congestion as the network scales, whereas the proposed quartic-coupling architecture embeds the interaction directly within localized oscillator coupling and eliminates the need for separate reference-comparison routing structures. Although horizontal and vertical coupling lines remain present, these interconnects are structured and localized, resulting in substantially lower routing overhead and improved physical scalability, as further confirmed by the post-layout ASIC implementation results shown 
in Figure~\ref{fig:Hardware_and_layout}b and Figure~\ref{fig:Hardware_and_layout}c.

\section{Related Work} \label{sec:Related Work}

In recent years, optimization using higher-order Ising machines has gained attention for solving complex combinatorial problems. The work presented in ~\cite{Bybee2023} investigates the use of Hopf oscillators and their amplitude dynamics, specifically mapping the SAT problem into a constraint satisfaction framework grounded in energy states. Here, computing with higher-order interactions requires a state variable to form and accumulate the partial derivatives of all terms in the total energy it participates in. The accumulation of partial derivatives can be a communication bottleneck eventually, and unlike our approach, they do not provide comprehensive hardware compatible solutions for their implementation.

Another article,~\cite{Bashar2023} explores the use of higher-order spin interactions in OIMs for solving combinatorial optimization problems such as NAE-k-SAT and max-k-cut. The authors formulate a dynamical system and the corresponding energy function based on phase dynamics, enabling the encoding of complex optimization problems. However, this implementation has a potential challenge with the system's tendency to become trapped in local minima. As the size of the problem graph increases, the role of these local minima in the high-dimensional phase space becomes more significant, posing potential limitations on the system's ability to efficiently solve larger-scale problems. Our implementation avoids this problem by using programmable couplers instead of altering the phase dynamics.

In this line of research,~\cite{Sharma2023} investigated the enhancement of cubic interactions in the Bistable Resistively-Coupled Ising Machine (BRIM) through additional architectural support,~\cite{Nikhar2024} implemented third-order interactions in a large-scale FPGA-based p-computer using hypergraph coloring to achieve parallelism,~\cite{10210653} proposed a reconfigurable higher-order Ising machine that utilizes SRAM to store spin interaction coefficients and employs a multiply-and-accumulate (MAC) unit for spin operations,~\cite{Hayato2023simulated} extended simulated bifurcation for higher order cost functions. However, none of these solutions cater to OIMs as opposed to ours and the operation cost is higher as these implementations rely on {multi-cycle spin operations}  which is not feasible for larger problems.

\section{Conclusion}\label{sec:Conclusion}

{In this paper, we propose a practical higher-order coupler to enable  Higher Order Oscillatory Ising Machines (HOIM) which can directly support interactions between an arbitrary number of problem variables.
We quantitatively compare the performance to state-of-the-art Oscillatory Ising Machines (OIM) which can only support pairwise interactions, using representative combinatorial optimization problems. 
We demonstrate how enabling higher-order coupling can reduce time- and energy-to-solution significantly.
HOIM, by construction, lifts the pressure on problem decomposition and reduces decomposition-induced adverse
effects on solver performance.
At the same time, by simplifying problem translation,
HOIM can reduce the likelihood of excessive scaling and
rounding of interaction coefficients, which can also degrade solver performance.


The key contribution of our paper is a practical higher-order coupler design, which can enable seamless problem translation for any even number of variable interaction,
by eliminating the reliance on external reference spins and by using a simple, parity-based logic.
Importantly, our coupler's ability to model all lower-order interactions further enhances its versatility.
{The core principle behind our quartic coupler design naturally extends to any even number of interactions beyond four.}

Apart from k-SAT problems, there exist other classes of COPs -- such as Max-k-Cut and hypergraph coloring -- that incorporate higher-order problem variable interactions. As the core principle behind our coupler design
is not limited to any specific problem formulation,
we can also 
support these and similar problems.


\bibliography{Reference}

\section*{Author Contributions}
 N. S. P. (corresponding author) was responsible for the development of the design and conducted the experiments. H. C. assisted with the formulation. A. K. and A.E. assisted with the software implementations. S. S. and U. K.  led the application mapping and benchmarking efforts.


\section*{Acknowledgments}
\textbf{Funding:} 
This work was supported in part by the Defense Advanced Research Projects Agency (DARPA) Quantum-Inspired Classical Computing (QuICC) program under Air Force Research Laboratory (AFRL) contract FA8750-22-C-1034, National Science Foundation (NSF) under Grants 2230963 and 2142248.

\section*{Data availability statement}
The datasets used and/or analyzed during the current study are available from the corresponding author on reasonable request.

\end{document}